\documentclass[sigconf]{acmart}

\usepackage{booktabs} 

\setcopyright{rightsretained}

\acmDOI{}

\acmISBN{}

\acmConference{Unpublished}
\acmYear{2021}
\copyrightyear{2021}

\acmArticle{}
\acmPrice{}

\editor{}
\editor{}
\editor{}

\setcopyright{none}
\begin{document}
\title{The Need for a Fine-grained approach in Just-in-Time Defect Prediction}

\author{Giuseppe Ng}
\orcid{0000-0002-1375-0236}
\affiliation{%
  \institution{De La Salle University}
  \streetaddress{2401 Taft Ave.}
  \city{Malate}
  \state{Manila}
  \postcode{1004}
}
\email{giuseppe_ng@dlsu.edu.ph}

\author{Charibeth Cheng}
\affiliation{%
  \institution{De La Salle University}
  \streetaddress{2401 Taft Ave.}
  \city{Malate}
  \state{Manila}
  \postcode{1004}
}
\email{charibeth.cheng@dlsu.edu.ph}

\begin{abstract}
With software system complexity leading to the rise of software defects, research efforts have been done on techniques towards predicting software defects and Just-in-time (JIT) defect prediction which predicts whether a code change is defective. While using features to determine potentially defective code change, inspection effort is still significant. As code change can impact several files, we investigate an open source project to identify potential gaps with features in JIT perspective. In addition, with a lack of publicly available JIT dataset that link the features with actual commits, we also present a new dataset that can be utilized in JIT and semantic analysis.
\end{abstract}

%
%



\maketitle

\section{Introduction}
Risks in software defect have increased over time  \cite{Ge2018,Huang2018,Ghose2018,Li2018,Nam2018,Hoang2019}. For instance, the Heartbleed bug was due to an OpenSSL vulnerability \cite{Hoang2019}. This vulnerability affected servers that utilized the library for secure connections such as Apache and Nginx, allowing an attacker to remotely read vulnerable server protected memory and potentially retrieve cryptographic information. The vulnerability affected an estimated 24-55\% of popular HTTPS websites \cite{Durumeric2014}. Such risks had lead towards the development of Software Defect Prediction (SDP) that predicts whether specific files in a software project are buggy or clean \cite{Ghose2018,Li2018,Dam2019,Son2019, Qiao2019,Hoang2019,Hosseini2019}. Finding software defects remains to be extremely challenging as tools could yield false positives predictions \cite{Ghose2018,Rahman2020}. In addition, this technique, however, is coarse granularity level of prediction and focuses on files or packages that are defective \cite{Yang2015,Albahli2019}.

Mockus and Weiss proposed changing the focus from files and packages to submitted code changes into VCS. This technique is referred to as Just-in-Time Defect Prediction (JIT) \cite{Mockus2000,Huang2017,82ac60c1c56b4be4a6220de0dfad6039}. Several works were done in the field of JIT and the crafting of features towards better performance \cite{Kamei2013,Yang2016,Fu2017,Qiao2019,Hoang2019}. 

\begin{figure}[h]
        \centering
        \includegraphics[width=\linewidth]
        {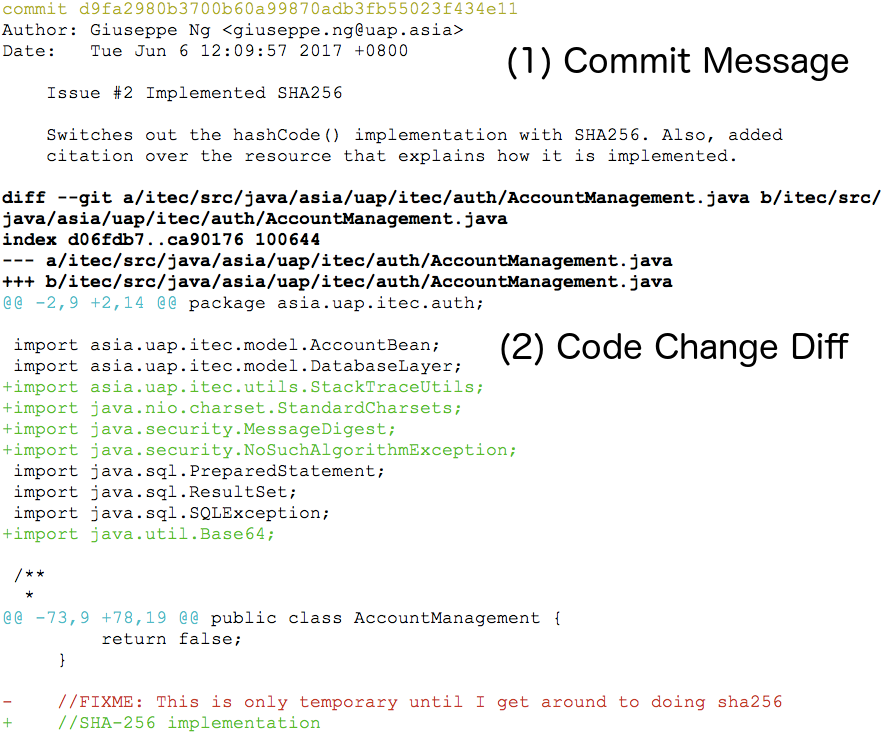}
        \caption{\label{fig:deepjit-diff} Sample code commit.}
    \end{figure}

When developers submit code changes to the repository, the developer submits a message regarding the change. The repository then generates the code diff \cite{Hoang2019}. A typical submitted code change can be seen in figure \ref{fig:deepjit-diff}.

The sample code commit on figure \ref{fig:deepjit-diff} shows two basic sections. The first section pertains to information about the commit, such as author, the timestamp, and a brief description of the submission. The other section shows the changes in the code compared to the previous version. These are denoted by the file name and path, the line numbers, the '+' symbol for line additions and '-' symbol for line deletions \cite{Hoang2019}.

Determining defective classes or files has the downside of lack of explainability of its defectiveness, leading to the inspection of code change as an alternate approach \cite{Yang2016,Huang2018,Dam2019}. Building datasets is a major aspect in machine learning approaches \cite{Ozaknc2018,Catal2009,Rahman2020}.

SDP datasets are based on analysis of Version Control Systems (VCS) such as GIT, SVN, or CVS, with defect information extracted from bug ticketing system  \cite{Li2018,Hosseini2019,Liu2019,Qiao2019}. Various traditional SDP datasets were made available online to aid in empirical software engineering and software defects prediction. These datasets were created using open source and closed source projects  \cite{Shepperd2013,Lamkanfi2013,Kamei2013,msr-2019-swh}. The ReLink dataset uses the following projects: (1) Apache, (2) Safe, and (3) ZXing to predict whether a source code file is defective \cite{Wu2011}. The AEEEM dataset is composed of the following open source projects: (1) Equinox (EQ), (2) JDT Core, (3) Lucene (LC), (4) Mylyn (ML), and (5) PDE UI. This dataset is intended to predict whether a class is defective or not \cite{DAmbros2012}. MORPH utilized several open source projects such as: ant-1.3, arc, camel-1.0, poi-1.5, redaktor, skarbonka, tomcat, velocity-1.4, xalan-2.4, and xerces-1.2 to predict whether a class is defective  \cite{Peters2012}. Shepperd et al. released the NASA dataset that utilized closed source projects inside NASA for traditional software defect prediction \cite{Shepperd2013}. SOFTLAB also used closed source projects for their dataset\cite{Turhan2009,Nam2018}. The NASA and SOFTLAB datasets predicts defects on a function level \cite{Shepperd2013, Turhan2009, Nam2018}.

When creating a dataset, in general, once the corpus has been identified, the predictor metrics need to be calculated \cite{Canfora2015}. The selection of traditional software metrics depends largely on the context and purpose defined \cite{Voas2017}. In terms of context of traditional SDP, metrics describing a software's size, complexity, and quality are critical \cite{Kan:2002:MMS:559784}. CK metrics \cite{Chidamber1994} and Lines of Code (LOC) are possible metrics that give different insight on the source code data \cite{Li2018,Hosseini2019}. CK metrics measures the complexity such as number of subclasses, depth of inheritance, number of methods to determine level of complexity \cite{Chidamber1994,Li2018}.  Nagappan and Bell showed that code churn, the measurement of a file changes over time, could be used as a feature \cite{Nagappan2005}. Scattered changes in the code could lead to defective software \cite{Hassan2009}. 

Process metrics could be used as well in determining defective code \cite{Moser2008}. Process metrics pertains to the use of information from source tracking repositories such as version control and issue tracking systems to capture development changes over time. These can be classified as: (1) code change, (2) developer-related metrics such as number of engineers, (3) code dependency, (4) organization-metrics such as organization volatility, (5) other process metrics \cite{Li2018}.

The work by \cite{Kamei2013} builds from the findings of early works and provided a dataset that included several projects for JIT defect prediction. This dataset has been used in several JIT works \cite{Kamei2013,Yang2016,Fu2017,Qiao2019,Hoang2019}.  The projects utilized for this are Bugzilla (BUG), Columba (COL), Eclipse JDT (JDT), Eclipse Platform (PLA), Mozilla (MOZ), and PostgreSQL (POS). The Bugzilla and Mozilla  datasets were created from data provided by the Mining Software Repository 2007 Challenge\footnote{\href{http://2007.msrconf.org/challenge/}{http://2007.msrconf.org/challenge/}}. The Eclipse JDT and Platform data were created from the MSR 2008 Challenge data\footnote{\href{http://2008.msrconf.org/challenge/}{http://2008.msrconf.org/challenge/}}. Columba and Postgres were from their respective CVS repositories \cite{Kamei2013}.

As a change-level defect prediction or JIT dataset, the granularity here is neither class, file, nor function. Instead, the features extracted were based on code change or commits in a code repository as explained in figure \ref{fig:deepjit-diff}. The features extracted are grouped into several dimensions: (1) Diffusion, (2) Size of the change, (3) Fix metric, (4) Historical information regarding the change, and (5) Experience of developers \cite{Kamei2013,Fu2017}. The features range from number of modified systems, files and directories, entropy, lines added and deleted, etc  \cite{Kamei2013,Lewis2013,Yang2016,Qiao2019}. Labelling a commit as bug inducing is done through the SZZ algorithm \cite{Sliwerski2005,Kamei2013}. In the case of Columba and PostgreSQL, since the defect identifiers were not included in the changelog or commit messages, an approximate SZZ approach was done where keywords such as "fixed" or "bug" would indicate a bug fix for a defect inducing commit \cite{Kamei2013}.

A summary of the software metrics used for JIT are the following \cite{Kamei2013, Yang2016, Fu2017, Chen2018}:

\begin{enumerate}
\item Number of modified subsystems;
\item Number of modified directories;
\item Number of modified files;
\item Entropy or the distribution of the modified code across each file;
\item Lines added;
\item Lines removed;
\item Lines of code in a file before the current change;
\item Whether the change is a bug fix; 
\item Number of developers that changed the modified file; 
\item The average time interval between the last and the current change; 
\item The number of unique changes to the modified file; 
\item Developer experience in terms of number of changes; 
\item Recent developer experience; and
\item Developer experience on the subsystem \cite{Kamei2013}. \end{enumerate}

The work by \cite{Hoang2019} also provided a dataset consisting of two open source projects: (1) Qt\footnote{\href{http://www.qt.io/}{http://www.qt.io/}}, and (2) Openstack \footnote{\href{http://www.openstack.org/}{http://www.openstack.org/}}. These datasets were cleaned by McIntosh and Kamei \cite{Hoang2019,10.1145/3180155.3182514}. This dataset contained commit ids unlike other datasets that made it possible to identify defect inducing commits and look at the actual change log of such \cite{Hoang2019}. To our knowledge, there has been no other publicly available dataset that has this information. As such, it is clear that many JIT approaches rely on feature metrics such as \cite{Kamei2013, Yang2016, Fu2017, Chen2018,Qiao2019,Pascarella2019}.

As conventional JIT approaches rely on feature metrics for defect prediction, we investigate whether code change feature metrics are adequate. In addition, we analyze the nature of code changes to show that inspection effort is still significant.

\section{Background}
\textit{Software defects} are as deficiencies in elements of a software because they do not meet requirements and specifications. \textit{Software faults} are defined as the causes of the defect while \textit{software failure} is defined as the end result of the defect \cite{Bourque2014,Ge2018}. 

In terms of software quality, several factors are considered with regards to testing requirements beyond the functional requirements of software. Some of these are: domain specific, possibly based on external and internal components integrated into the system; and intended expertise of the end user \cite{Bourque2014}.

Factors that lead to software defects are: human-related, lack of communication, insufficient design and testing, unrealistic timeframe, and defects from third party components, etc. \cite{Ge2018}. A defect taxonomy can be used to effectively group different kinds of software defects. These are \cite{Felderer2015}:
\begin{enumerate}
\item requirements;
\item features and functionalities;
\item structural defects;
\item data;
\item implementation and coding;
\item integration;
\item system software architecture;
\item test definition and execution; and
\item unclassified defects \cite{Felderer2015}.
\end{enumerate}

\subsection{Software Defect Prediction}
The goal of software defect prediction, then, is to identify the faults in the code and code modifications, estimate the number of defects and identify areas of optimization\cite{Hosseini2019}. This technique falls under data-driven software engineering \cite{Upadhyaya2018}.

The learning models in SDP can generally be classified under supervised or non-supervised  \cite{Catal2009,Ge2018,Li2018,Ozaknc2018,Hosseini2019}. The model is expected to learn from a corpus of data that encompasses code archive (in classes or files form and source trees such as GIT, SVN, or CVS), and defect information (in a bug ticketing system such as Bugzilla or Jira) \cite{Li2018,Hosseini2019,Rahman2020}. Linking source commit information and bug tickets may also be necessary. Granularity or instance of defect prediction can be by package, class, file, or function basis \cite{Qiao2019,Hosseini2019}.

Training data can be from the same project or referred to as 'Within Project Defect Prediction' (WPDP) or from other projects, 'Cross Project Defect Prediction' (CPDP) \cite{Hosseini2019}. Traditional defect prediction models used WPDP project where older versions of the software are used to predict defects on newer versions \cite{Li2019}.

\begin{figure}[!htb]
        \center{\includegraphics[width=0.9\linewidth]
        {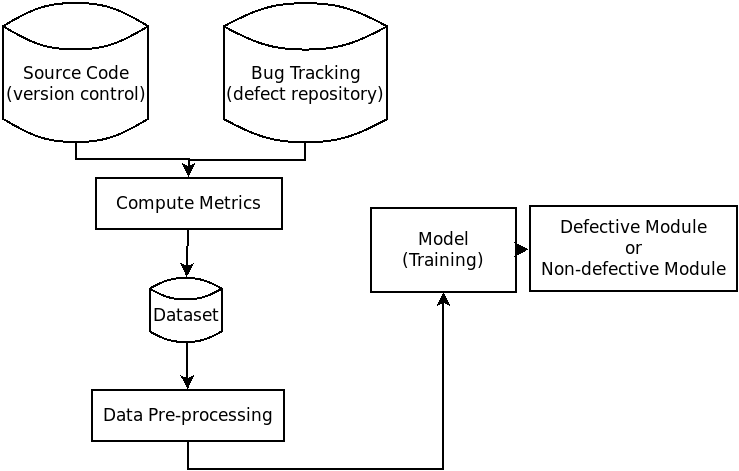}}
        \caption{\label{fig:model-design} Generic Flow of Designing a Default Prediction Model}
    \end{figure}

The work by \cite{Canfora2015} outlines the basic workflow in designing a prediction model. This can be seen in figure \ref{fig:model-design}.

The traditional process in building the model involves extracting the features, and building the ML model that will be trained with this model. Given the instance whether it's a package, a file, a class, or a method, the ML model should be able to classify the instance as defective \citep{Yang2015,Albahli2019,Qiao2019,Rahman2020}.

\subsection{Just-in-Time Defect Prediction}
\begin{figure}[!htb]
        \center{\includegraphics[width=\linewidth]
        {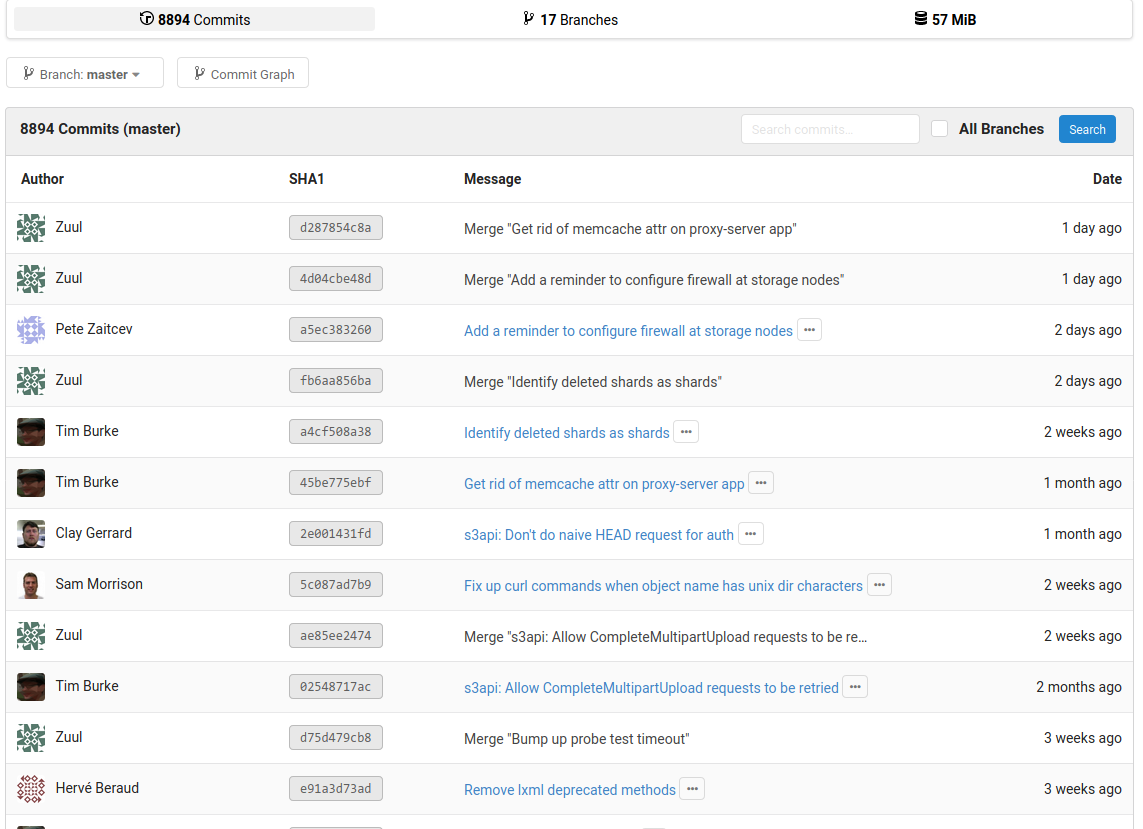}}
        \caption{\label{fig:openstack-git} Repository of Commits from Openstack.}
    \end{figure}

Where traditional SDP have been focused on the predicting the coarse-grained instances as defective, others have expanded SDP to analyzing code changes instead of files to identify whether the change is defective. This is referred to as Just-in-Time (JIT) defect prediction \cite{Hata2012,Fukushima2014}. Figure \ref{fig:openstack-git} illustrates a typical code repository\footnote{\href{https://opendev.org/openstack/swift/commits/branch/master}{https://opendev.org/openstack/swift/commits/branch/master}} where all the commits are stored.

\begin{figure}[!htb]
        \center{\includegraphics[width=\linewidth]
        {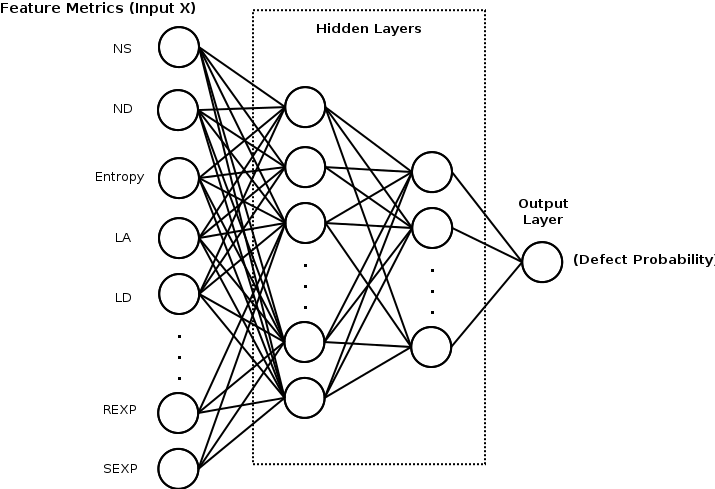}}
        \caption{\label{fig:qiao-jit} Neural Network Architecture, from \citep{Qiao2019}.}
    \end{figure}

For every code change in the code repository, features are extracted that are then used to train the model \cite{Kamei2013,Fu2017}. Figure \ref{fig:qiao-jit} shows the approach of a JIT defect prediction \cite{Qiao2019}.

Unlike traditional SDP, the granularity in JIT is actual code changes, achieving more fine-grained approach \cite{Cho2018}.

As discussed by Kamei \cite{Kamei2013}, the diffusion aspect is represented by NS, ND, NF and Entropy. Highly distributed changes are more complex and are more likely to be defective. Number of modified subsystems, modified directories, and modified files using the root directory name, directory names, and file names respectively \cite{Kamei2013}.

\begin{figure}[htbp]
\centerline{\includegraphics[width=0.5\textwidth]{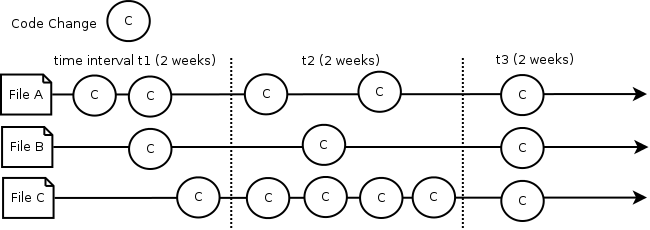}}
\caption{Example of Entropy \cite{DAmbros2010}.}
\label{fig:entropy}
\end{figure}

Entropy is calculated using a similar measure to Hassan \cite{Hassan2009,Kamei2013}. It is the measurement over time how distributed the changes are in the code base. Changes in one file is simpler than one that impacts many different files. Time period used for the dataset is 2 weeks \cite{Kamei2013}. Entropy is defined as \cite{Kamei2013}:

\begin{equation}
H(P)=- \sum_{k=1}^{n}(p_k*log_2 p_k) \label{eq:entropy}
\end{equation}

where probabilities $p_k \geq 0, \forall k \in 1,2, ..., n$, $n$ is the number of files in the change, $P$ is a set of $p_k$, where $p_k$ is the proportion that $file_k$ is modified in a change and $(\sum_{k=1}^{n} p_k) = 1$ \cite{Kamei2013}.  
Figure \ref{fig:entropy} illustrates entropy in a sample scenario \cite{DAmbros2010}.

For time interval $t1$, there are four changes and the probabilities are $P_A = \frac{2}{4}, P_B = \frac{1}{4}, P_C = \frac{1}{4}$. The entropy in $t1$ is calculated as $H = -(0.5 * log_2 0.5 + 0.25 * log_2 0.25 + 0.25 * log_2 0.25) = 1$. For time interval $t2$, entropy is 1.378 \cite{DAmbros2010}.

Lines added (LA), lines deleted (LD), and lines of code before the current change (LT) were calculated directly from the source code. LA and LD were normalized by dividing by LT while LT is normalized by dividing by NF since these two metrics have high correlation \cite{Kamei2013}. NUC is calculated by counting the number of commits that caused changes to specific files. This metric was also normalized by dividing by NF due to their correlation with the NF feature \cite{Kamei2013}. This technique follows findings made by Nagappan and Ball \cite{Nagappan2005}.

Age is calculated as the average of time interval between the last and current change of files \cite{Fu2017,Kamei2013}. Should Files A, B and C were modified 3 days ago, 5 days ago, and 4 days ago respectively, Age is calculated as 4 (i.e., $\frac{3+5+4}{3}$). More detailed explanation of the features of the dataset can be found in \cite{Kamei2013}.

Experience is the number of commits done by the author prior to the current commit while REXP is the total experience of the developer while factoring the recency of the change. Using $\frac{1}{n+1}$ where $n$ represents the year, a developer who made 5 changes 3 years ago would be represented as $\frac{5}{3}$. REXP is then computed as the summation of all such factored EXP values \cite{Kamei2013,Borg2019}.

\subsection{SZZ Algorithm}
Labeling defective changes is done through the SZZ algorithm \cite{Sliwerski2005}. Two key concepts in labeling is defining a bug inducing change and a bug fixing change. Bug inducing change is defined as a change A that caused for a change B to be submitted to rectify an issue introduced by change A. Change B is referred to as the bug fixing change \cite{Kiehn2019}. An open source implementation was created by \cite{Borg2019}. With regards to the Columba and Postgres dataset, an approximated SZZ approach was taken due to the defect ids not referenced in the commit messages \cite{Kamei2013}.
 
The overall workflow of the algorithm is as follows \cite{Borg2019}:
\begin{enumerate}
	\item Retrieve all closed bug reports from a ticket tracker of the system
	\item Link the bug report to the bug fixing commits of the code repository
	\item Identify the bug inducing commits through the bug fix commit information.
\end{enumerate}
 
\begin{figure}[!htb]
        \center{\includegraphics[width=\linewidth]
        {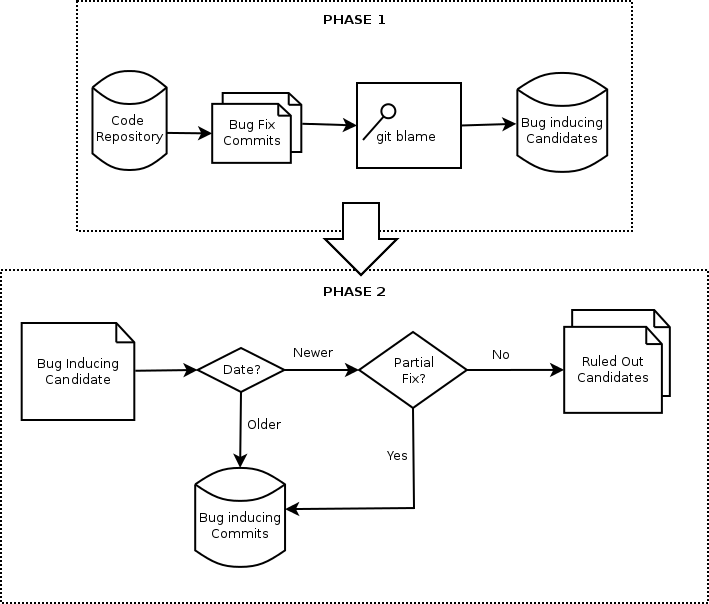}}
        \caption{\label{fig:szz} SZZ Phases \cite{Borg2019}.}
    \end{figure}

Phase 1 involves the linking of the bug reports to bug fixing commits while Phase 2 is the process of identifying the bug inducing commit. Phase 1 largely involves matching the commit to the closed ticket IDs \cite{Borg2019}. In the even that there is no ticket to match, words such as \textit{fix} in the commit message are assumed to be bug fixes \cite{Kamei2013,Borg2019}. Phase 2 is illustrated in Figure \ref{fig:szz} \cite{Borg2019}.

The tool will identify all commits that happened for specific lines using \textit{git blame}. Each candidate is then ruled out by checking the date of the commit. If the change is newer but it was only a partial fix, then it is included as a bug inducing commit \cite{Borg2019}.

The end result of the algorithm is the list of submitted commits fixing bugs and potential list of changes as bug inducing. Using the SZZ tool provided by \cite{Borg2019}, pairs of bug inducing commits and bug fix commits can be identified from a dump of the bug tracking and report system and the git repository. The commits that are not labelled as defective from SZZ can be used to build the non-defective labelled dataset \cite{Borg2019}.

Subtle semantic differences would be difficult to identify from calculated metrics \cite{Ghose2018}. In JIT, there are limitations in feature metrics from code changes as two different code changes can lead to the same code metrics, thereby highlighting the need to extract features from semantic code \cite{Hoang2019}.

\subsection{Semantic Representation of Code Change}
\begin{figure}[h]
        \centering
        \includegraphics[width=\linewidth]
        {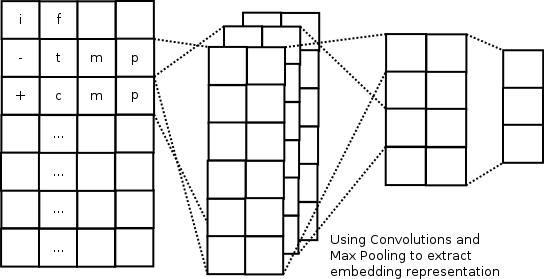}
        \caption{\label{fig:hoang-diff} Feature extraction of commit diff using CNN, from \cite{Hoang2019}.}
    \end{figure}

Although extraction of features from source code has been done for various purposes including SDP, the study of semantic features in JIT context is under-represented. \cite{Hoang2019} explored extracting features from semantic information of a code change.  In their work, the code change, represented by the set of lines added and deleted, and the commit message were used in feature extraction. Figure \ref{fig:hoang-diff} illustrates their approach \cite{Hoang2019}.

Their work shows that utilizing the semantic information from code changes can lead towards JIT \cite{Hoang2019}. 

\subsection{Comparing Coarse-grained and Fine-grained JIT}
\begin{figure}[!htb]
        \center{\includegraphics[width=\linewidth]
        {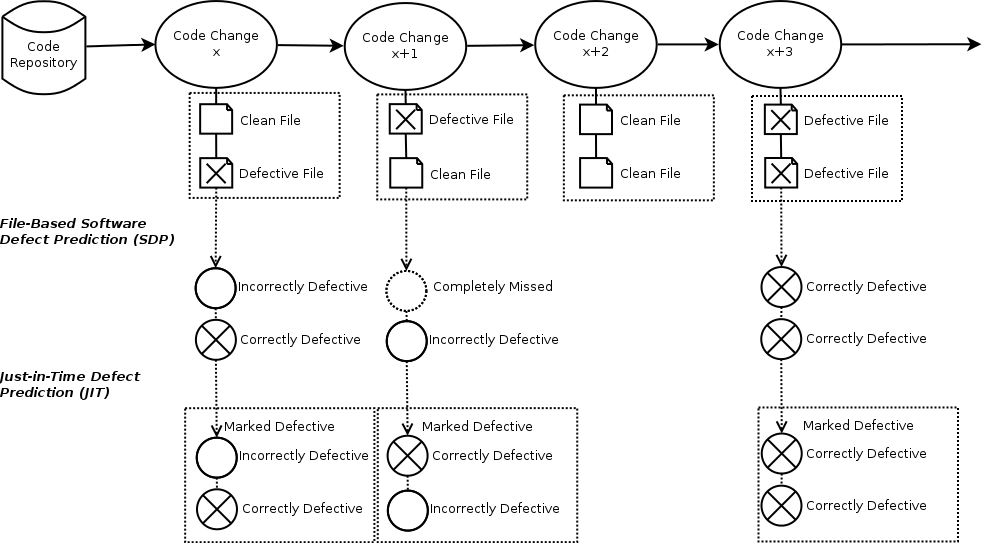}}
        \caption{\label{fig:pascarella-partialdefect} Partial defect and existing approaches, from \citep{Pascarella2019}.}
    \end{figure}

A commit or code change can either be clean, completely bug inducing, or partially bug inducing. Traditional SDP where the model used for prediction is based on previous release data. In this scenario, the predictor can incorrectly label a file as defective. It can also completely miss defect inducing files \citep{Pascarella2019}. Such an approach lacks feedback that developers need when doing code review \citep{Kamei2013,Lewis2013,Pascarella2019}. Figure \ref{fig:pascarella-partialdefect} illustrates how existing approaches are not adequate in identifying defective files within commits \cite{Pascarella2019}.

Regarding JIT defect prediction, coarse-grained JIT can predict whether an incoming code change is defect inducing. This approach, unlike the traditional SDP, can catch defect inducing changes as soon as it is made, thereby giving faster feedback to developers. However, the code inspection effort is still significant \cite{Lewis2013,Pascarella2019}. 

A submitted code change can modify more than one file and can lead to significant inspection cost by developers to determine what particular area of the code change was defective. In addition, not all the files in the code change are necessarily defect inducing, and yet, such approaches could cause defect-free files in the commit to be defect inducing \cite{Pascarella2019}. 

\begin{figure}[!htb]
        \center{\includegraphics[width=\linewidth]
        {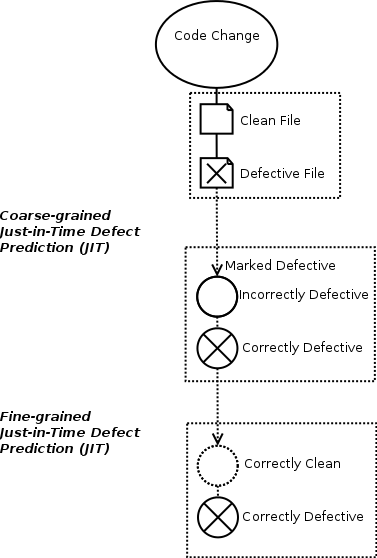}}
        \caption{\label{fig:pascarella-coarsefine} Difference of Coarse vs Fine-grained}
    \end{figure}

Hence, a finer grained approach towards JIT that is able to analyze changes within a code change could reduce the inspection cost. Ideally, this would entail identifying actual files that are defective and even the area within that file in the code change to be inspected. Figure \ref{fig:pascarella-coarsefine} illustrates the difference of perspective of coarse-grained and fine-grained JIT.

As shown, the coarse-grained JIT can label a change as defective but in doing so, incorrectly labels a file as defective when it is in fact clean.  A finer-grained JIT can correctly mark the actual file within the code change as defective and leave the clean file as such. Efforts to improve fine grained defect prediction echo the findings by \cite{Hata2012,Pascarella2018,Liang2019,Pascarella2019}.

\section{Methodology}
To determine whether feature metrics can determine the nature of defective code change, our first step is to create a dataset using an open source project as code repository and code change data are available. To build the dataset, we downloaded the source code repository and exported the defect data from an open source project. This is consistent with the work done by \cite{Kamei2013,Borg2019}. The inclusion of the commit id in the dataset is consistent with the work done by \cite{Hoang2019}. We then investigated the marked bug inducing commits of the project. Specifically, we examine the nature of code changes and report our observations in relation to the calculated feature metrics.

To see the predictive performance of the dataset, we use a Within-Project Defect prediction approach, a practice that Google used with known existing models \cite{Kim2007,Rahman2011,Rahman2012}.

To validate our findings, we referred to the dataset provided by \cite{Hoang2019} to see if the bug inducing commits in the dataset are consistent.

\section{Creating and Labelling a Dataset}
We downloaded the git source repository of the \textit{Trac} project\footnote{\href{https://github.com/edgewall/trac}{https://github.com/edgewall/trac}} and also obtained an export of their list of bugs\footnote{\href{https://trac.edgewall.org/query}{https://trac.edgewall.org/query}}. The export covers tickets from August 24, 2003 to January 13, 2020.

Using the SZZ tool provided by \cite{Borg2019}, pairs of bug inducing commits and bug fix commits can be identified from a dump of the bug tracking and report system and the git repository. The software metrics from the project was calculated using the scripts provided. The commits are labelled accordingly using the SZZ tool by \cite{Borg2019}.

There are 12,840 commits processed in the repository. The SZZ tool labelled 1681 commits as bug inducing. 1552 bug fixes were found by the tool. The information is shown in table \ref{tab:tracdatasetsummary}.

\begin{table}[htbp]
\caption{Summary of the information on the Trac dataset}
\begin{center}
\begin{tabular}{c|c|c|c}
 &  &  & \textbf{\%Defect} \\
 &  &  & \textbf{induced} \\
\textbf{Project} & \textbf{Period} & \textbf{\#change} & \textbf{change} \\
\hline
Trac & 08/2003-01/2020 & 12840 & 13.09\% \\
\end{tabular}
\label{tab:tracdatasetsummary}
\end{center}
\end{table}

We used \textit{cloc} software\footnote{\url{http://cloc.sourceforge.net}} to calculate the lines of code (LOC) of the project. This is used to determine the size of the project being processed. Table \ref{tab:trac-cloc} illustrates the statistics of the code.

\begin{table}[htbp]
\caption{CLOC Summary of the Trac Project}
\begin{center}
\begin{tabular}{c|c|c|c|c}
 \textbf{Language} & \textbf{files} & \textbf{blank} & \textbf{comment} & \textbf{code} \\
\hline
Python & 323 & 16624 & 23054 & 74407 \\
HTML & 73 & 1126 & 32 & 9984 \\
CSS & 17 & 294 & 185 & 4206 \\
Javascript & 25 & 261 & 327 & 2346 \\
DTD & 1 & 200 & 214 & 564 \\
make & 1 & 201 & 93 & 557 \\
PowerShell & 1 & 98 & 114 & 209 \\
YAML & 2 & 7 & 5 & 207 \\
CoffeeScript & 2 & 22 & 50 & 93 \\
DOS Batch & 1 & 16 & 0 & 79 \\
Bourne Shell & 1 & 13 & 179 & 59 \\
Bourne Again Shell & 2 & 4 & 16 & 37 \\
XML & 1 & 0 & 0 & 12 \\
\hline
\textbf{Total} & \textbf{450} & \textbf{18866} & \textbf{24269} & \textbf{92760} \\
\end{tabular}
\label{tab:trac-cloc}
\end{center}
\end{table}

With this information we had managed to extract the features shown on table \ref{tab:tracdatasettable}.

\begin{table}[h!]
\caption{Features of the Trac dataset}
\begin{center}
\begin{tabular}{c|l}
\textbf{Metric} & \textbf{\textit{Description}} \\
\hline
Commit & The commit hash in the \\
& repository \\
\hline
NS & Number of modified subsystems \\
ND & Number of modified directories  \\
Entropy & Distribution of the modified \\
& code across each file \\
\hline
Lines\_of\_Code\_Added & Lines added  \\
Lines\_of\_Code\_Deleted & Lines deleted  \\
Files\_Churned & Number of modified files \\
Lines\_of\_Code\_Old & Lines of code in a file before the \\
 & current change \\
\hline
Purpose (Fix) & Whether the change is a bug fix \\
\hline
Number\_of\_Authors & Number of developers  \\
(NDEV) & that changed the modified file \\
AGE & The average time interval \\
 & between the last and the \\
 & current change \\
Number of & The number of unique \\
Unique Changes & changes to the modified file \\
\hline
EXP & Developer experience in terms of \\
& number of changes \\
REXP & Recent developer experience \\
\hline
Defective & Whether this is bug inducing \\
& or not \\
\end{tabular}
\label{tab:tracdatasettable}
\end{center}
\end{table}

There are some deviations from the feature extraction methods described in \cite{Kamei2013}. We highlight the notable differences in the calculations in the tool provided in \cite{Borg2019}. 

Entropy, Experience, and REXP are calculated similarly with Kamei \cite{Kamei2013}. Lines of code added and deleted are calculated by extracting the number of lines inserted and deleted respectively and dividing them by the total lines of code of the new file \cite{Borg2019}. Lines of code of the old file is not normalized unlike LT feature in Kamei's dataset \cite{Kamei2013,Borg2019}. 

Age is calculated as an average of the time difference of every file modified by the current commit. Number of unique changes is calculated as the number of commits for every file modified by the commit \cite{Borg2019}.

For the normalization on the dataset, we employ the min-max standardization from \cite{Qiao2019}:
\begin{equation}
	Norm(x) = \frac{x - min(x)}{max(x)-min(x)}
\end{equation}

This has been a commonly used technique to treat datasets in other works \cite{Wang2016,Qiao2019}.

\subsection{The Trac Dataset}
In this section, we report on interesting observations that we had seen from the dataset.

\textbf{Commit ID}. Our key inclusion in this dataset is the Commit ID for every submitted code change of the dataset. With this, it is possible to look into each and every change that was marked as bug inducing for further study. Many JIT datasets do not include this, making the change level semantic learning on change-level defect prediction much more challenging.

\textbf{Lines of Code Added and Deleted}. We compare the lines of code added and deleted and found an interesting difference between the maximum lines of code added and deleted. The maximum lines of code added is 1175 while the lines of code removed is 468343. In analyzing the increase of lines of code removed, we noted that there is a commit that removed almost all the files in the repository as a repository clean up. The presence of this submission is an outlier in the repository.

\textbf{Files Churned}. We  observed that since the tool already normalizes the number of modified files by the total number of files, the values are already in the range of 0 to 1. There are 39 instances when the values are not normalized due to potential divide by 0 issues that could happen when the submitted code change cleans up the repository. The minimum and maximum values of the feature are 0 and 739 respectively with an average of 0.1175 and a standard deviation of 6.5236.

\begin{figure}[!htb]
        \center{\includegraphics[width=\linewidth]
        {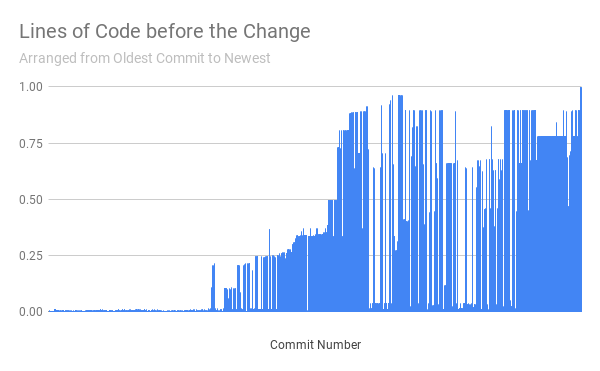}}
        \caption{\label{fig:lt-norm} Lines of Code before the Change Normalized.}
    \end{figure}

\textbf{Lines of Code before the Change (LT)}. The feature is not normalized and has an average of 141.7006 with a standard deviation of 326.9681. We found that over time, there is a general increasing trend in this value as shown in figure \ref{fig:lt-norm}.

\begin{figure}[!htb]
        \center{\includegraphics[width=\linewidth]
        {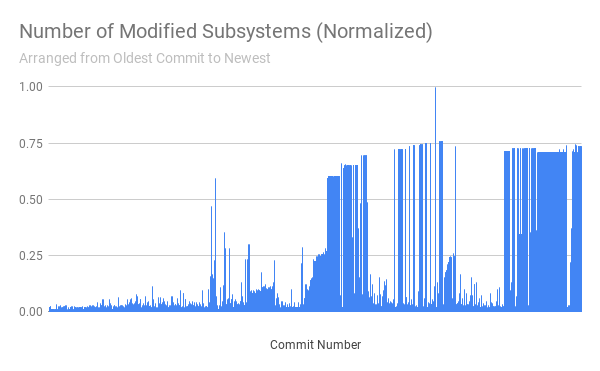}}
        \caption{\label{fig:ns-norm} Number of Modified Subsystems Normalized.}
    \end{figure}

\textbf{Number of Modified Subsystems}. Upon applying the normalization step, we observed that over time, there is an increasing trend in the values. Figure \ref{fig:ns-norm} illustrates this.

This would make sense that over time, the system complexity would grow as more features are added in. The minimum and maximum values of the feature are 0 and 198 respectively with an average of 0.1175 and a standard deviation of 33.9173.

\textbf{Number of Modified Directories}. The minimum and maximum values of the feature are 0 and 8 respectively with an average of 1.5326 and a standard deviation of 1.5488. These values are raw and not standardized.

\textbf{Purpose}. In identifying whether a code change is a bug fix, the feature is represented as a boolean value. In the entire 12,840 commits, the tool marked a total of 2318 bug fixes representing 18.05\% of all code commits to the repository.

\textbf{Entropy}. In terms of entropy, we found that the average entropy is 1.2246 with a standard deviation of 2.0082.

\begin{figure}[!htb]
        \center{\includegraphics[width=\linewidth]
        {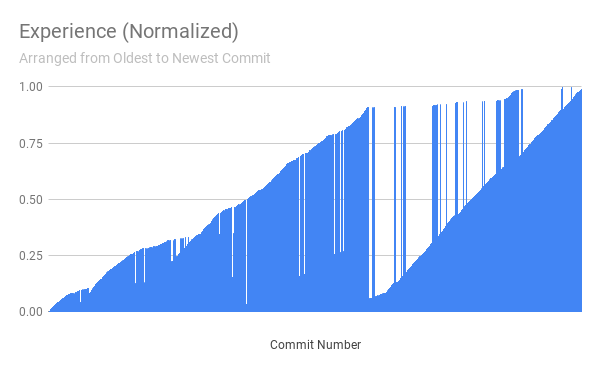}}
        \caption{\label{fig:exp-norm} Experience Normalized.}
    \end{figure}

\textbf{Experience and Recent Experience}. We found that the average is 1145.5416 with a standard deviation of 1014.5913. We found that the huge code clean up influenced the values of the feature by plotting the chart that is shown in figure \ref{fig:exp-norm}.

Analyzing the chart, shows an increasing trend on experience over time. However, because the code clean up removes many files, this also eliminates the experience that is tracked by looking at those specific files.

With the recent experience chart, the trend of the feature shows that the huge dip from the clean up does not influence the feature.

	\begin{figure}[!htb]
        \center{\includegraphics[width=\linewidth]
        {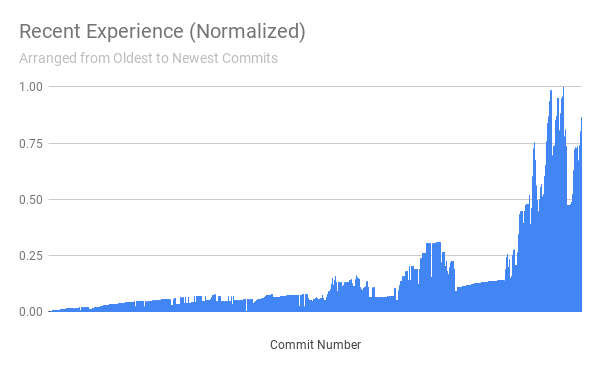}}
        \caption{\label{fig:rexp-norm} Recent Experience Normalized.}
    \end{figure}

\textbf{Number of Unique Changes (NUC).} The feature has an average of 2.0177 and a standard deviation of 3.9729. Interestingly, the maximum value is 180. This shows that this is an extreme outlier and suggests that standardization techniques may not be advisable.

\textbf{Defective.} This is the label of the dataset and marks a particular code change as defective.

\begin{figure}[!htb]
        \center{\includegraphics[width=\linewidth]
        {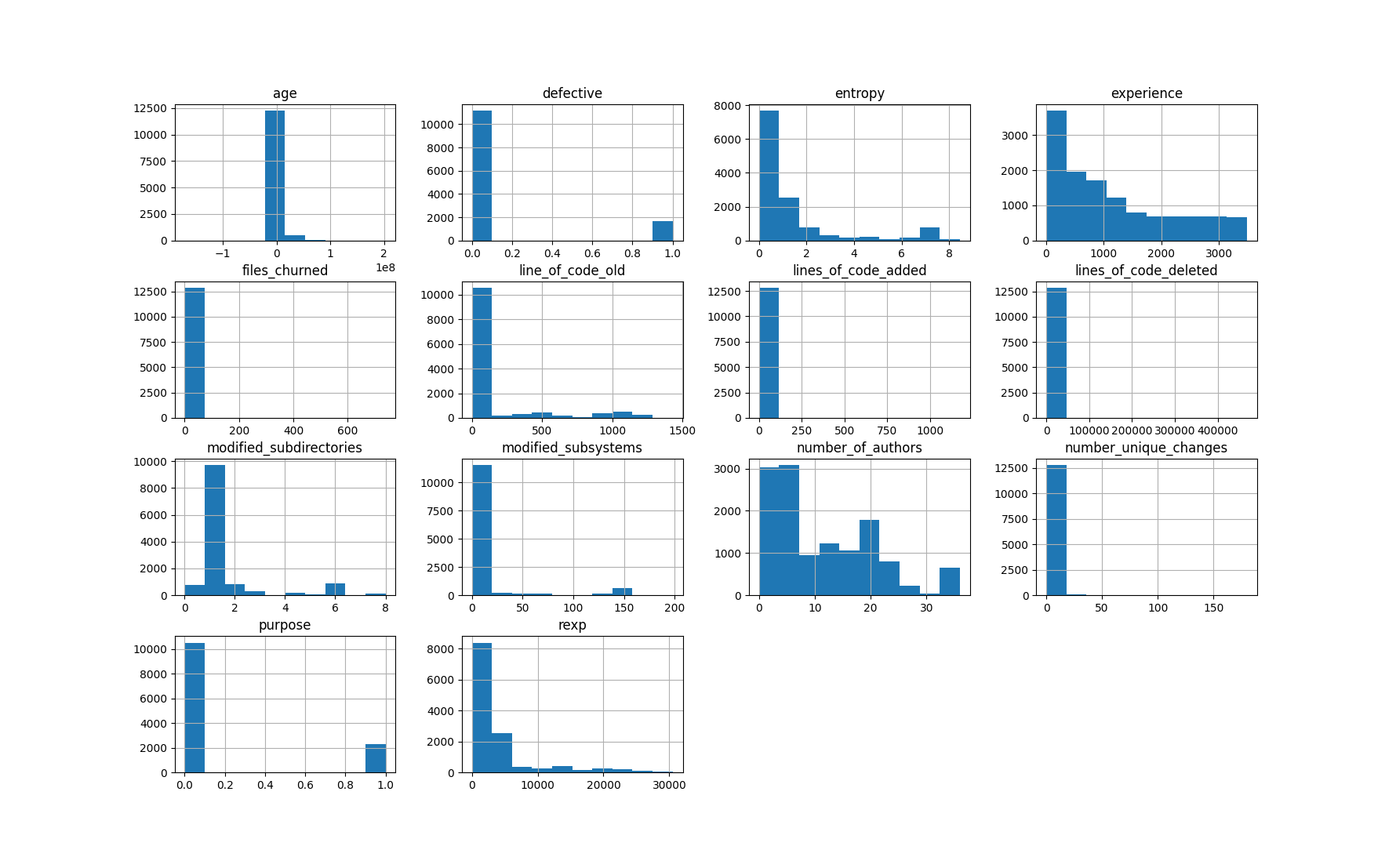}}
        \caption{\label{fig:histogram} Histogram of the Trac Dataset Features.}
    \end{figure}

\begin{figure}[!htb]
        \center{\includegraphics[width=\linewidth]
        {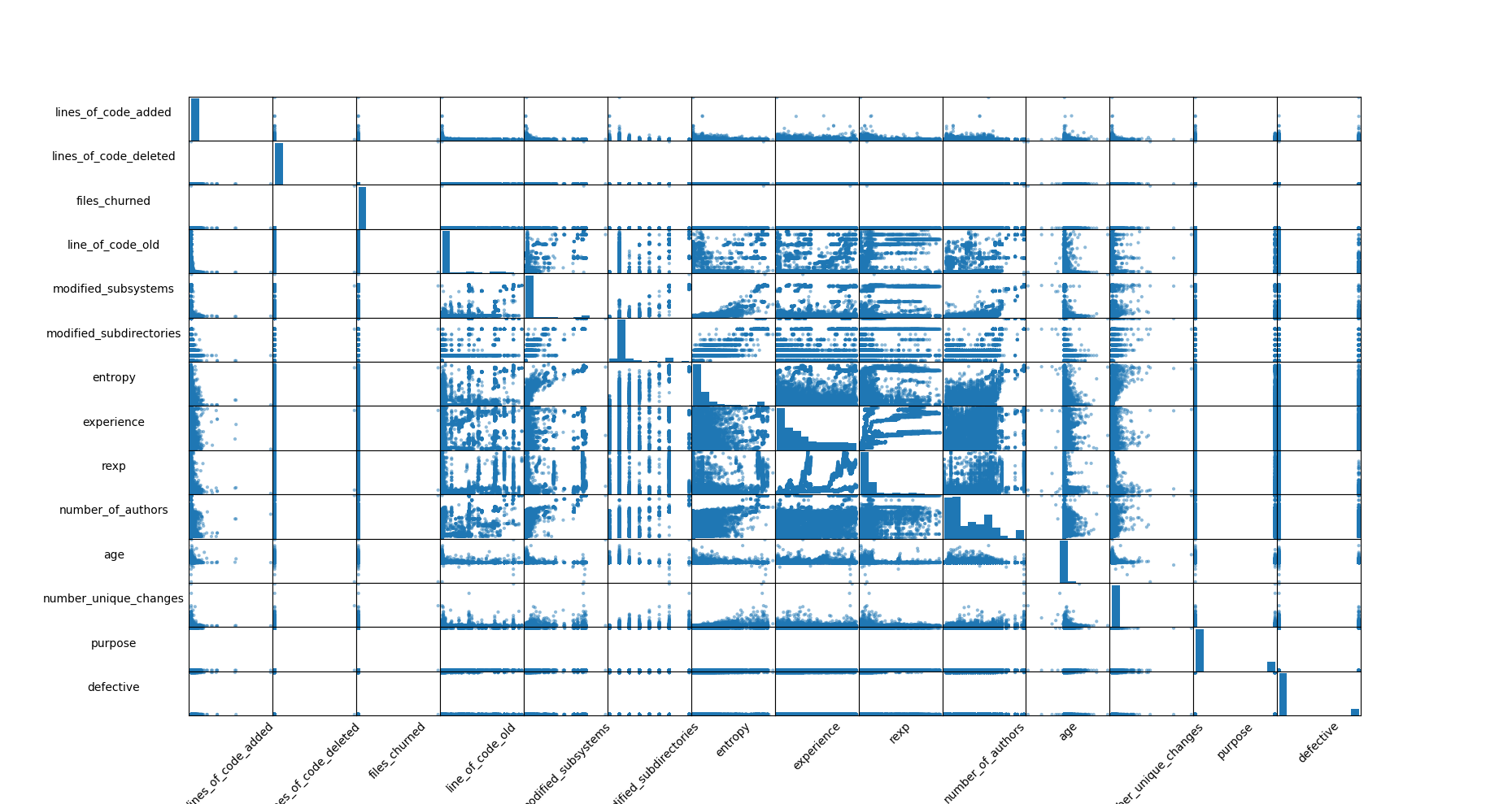}}
        \caption{\label{fig:scattermatrix} Scatter Matrix of the Trac Dataset Features.}
    \end{figure}

Figures \ref{fig:histogram} and \ref{fig:scattermatrix} show the histogram and the scatter matrix visualization of the dataset.

The statistics of the features of this dataset are summarized in table \ref{tab:tracstat}.

\begin{table}[h!]
\caption{Statistics of the Trac Dataset}
\begin{center}
\begin{tabular}{c|l|l|l}
\textbf{Feature} & \textbf{\textit{Min-Max}} & \textbf{\textit{Ave}} & \textbf{\textit{Std-Dev}} \\
\hline
LA & 0-1175 & 6.8046 & 20.4584 \\
LD & 0-468343 & 40.6886 & 4133.1486 \\
Files Churned & 0-739 & 0.1175 & 6.5236 \\
LT & 0-1429 & 141.7008 & 326.9681 \\
NS & 0-198 & 0.1175 & 33.9173 \\
ND & 0-8 & 1.5326 & 1.5488  \\
Entropy & 0-8.4429 & 1.2246 & 2.0082 \\
Exp & 0-3491 & 1145.5415 & 1014.5912 \\
NUC & 0-180 & 2.0177 & 3.9729 \\
\end{tabular}
\label{tab:tracstat}
\end{center}
\end{table}

In training an actual model to see its potential, we applied a 70-30 split of seen and unseen data into a 9 layer architecture of ReLU units and a sigmoid output. Our data preprocessing techniques were normalization and undersampling. We utilized Smooth L1 Loss as our loss function with 3500 epoch training. An Adam optimizer with a learning rate of 0.001 was used.

By repetitively removing features from the training and checking the loss and recall, we identified that \textit{lines of code added}, \textit{lines of code deleted}, \textit{modified subsystems}, \textit{modified subdirectories}, \textit{experience}, \textit{rexp}, \textit{age}, \textit{number unique changes} may be a potent starting set of features for training. 

The selection of \textit{modified subsystems}, \textit{age}, and \textit{number unique changes} as good features is in agreement with the results from \cite{Kamei2013}. 

The use of experience metrics such as \textit{experience} and \textit{rexp} is consistent with the findings from \cite{Matsumoto2010,Mockus2000,Pascarella2019} where experience-based metrics are good features.

Interestingly, our tests show that \textit{lines of code added}, \textit{lines of code deleted}, and \textit{modified subdirectories} may be included as well in the list of good features.  Our tests managed to attain a recall of 64.15\% with a loss of 0.038 in a WPDP approach.

\subsection{Observations of the Trac repository code changes}
We would like to make note that in this study, we do not claim to be experts of the Trac source project. The analysis below is purely based on the information we had extracted from the repository.

\begin{figure}[!htb]
        \center{\includegraphics[width=\linewidth]
        {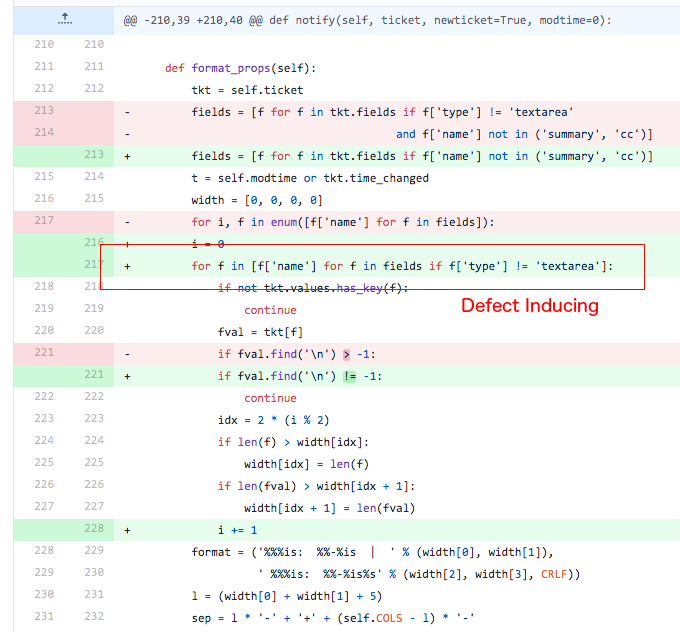}}
        \caption{\label{fig:defectinducingtrac} Relevant Snippet of Defect Inducing Change.}
    \end{figure}

In observing bug inducing changes and bug fixes, we make note that bug fixes tend to modify lines that were added by bug inducing changes. In doing code comparisons of defective code, figure \ref{fig:defectinducingtrac} shows a defect inducing code change from the Trac repository\footnote{Commit hash \texttt{f1ab6093ddcbd5043fbfaafca5b5e06bde87313b}}. 

\begin{figure}[!htb]
        \center{\includegraphics[width=\linewidth]
        {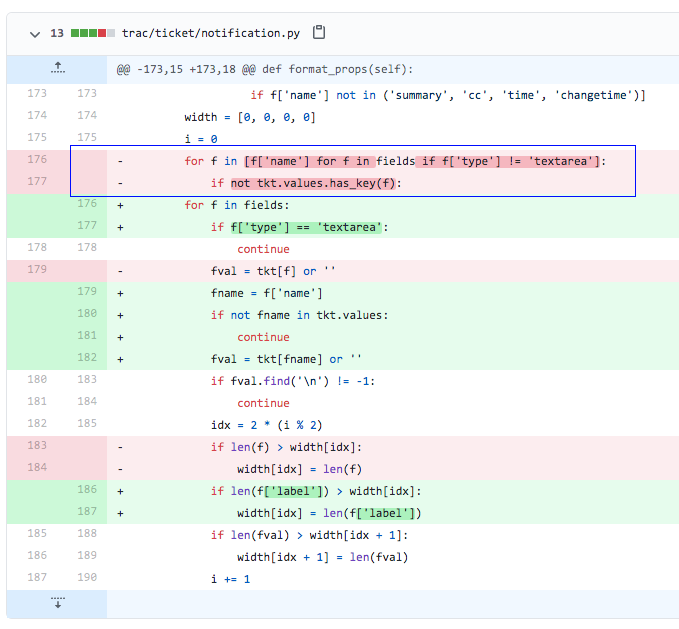}}
        \caption{\label{fig:bugfixtrac} Relevant Snippet of the Bug Fix.}
    \end{figure}

In contrast, a sample bug fix code change\footnote{Commit hash \texttt{936959307a5421c43ffa2b6024801f80c8b720c8}} is shown in figure \ref{fig:bugfixtrac}.

To identify defect inducing lines, we compare the bug inducing change and the bug repair change for lines that were modified. Lines that were modified from the bug inducing change to the bug fix are defect inducing, while lines that were untouched are not defect inducing. This is consistent with the works by \cite{Yan2020,Wattanakriengkrai2020}.

The defect inducing change here only changes 1 file with 14 line additions and 13 line deletions and is a bug fix to another reported bug. This code change was submitted October 8, 2005. The bug fix for this was submitted May 12, 2010, 10 years after the original code change was accepted in the system. Clearly, this defect inducing change had already been released into actual stable versions of Trac.

Notice also that figure \ref{fig:bugfixtrac} also deleted lines 183 and 184, replacing them with 186, and 187. The lines were also in figure \ref{fig:defectinducingtrac}, however, the change itself did not introduce these lines. Instead, it is introduced by another bug fix\footnote{Commit e955cc298b14d4c8585186dd50d2db31c6327444}  submitted to the repository. This implies that these lines were also defect-inducing from a different code change in the repository. The defect itself may have been caused by shifting context over time, possibly by code or by the dependent library environment or extrinsic bugs as defined by \cite{Rodriguez-Perez2020}. We make note that a single bug fix change can be linked to several defect inducing code changes. It is possible also that defect was introduced by the perceived understanding of a developer of the code as pointed out by \cite{Fakhoury2020}.

\begin{figure}[!htb]
        \center{\includegraphics[width=\linewidth]
        {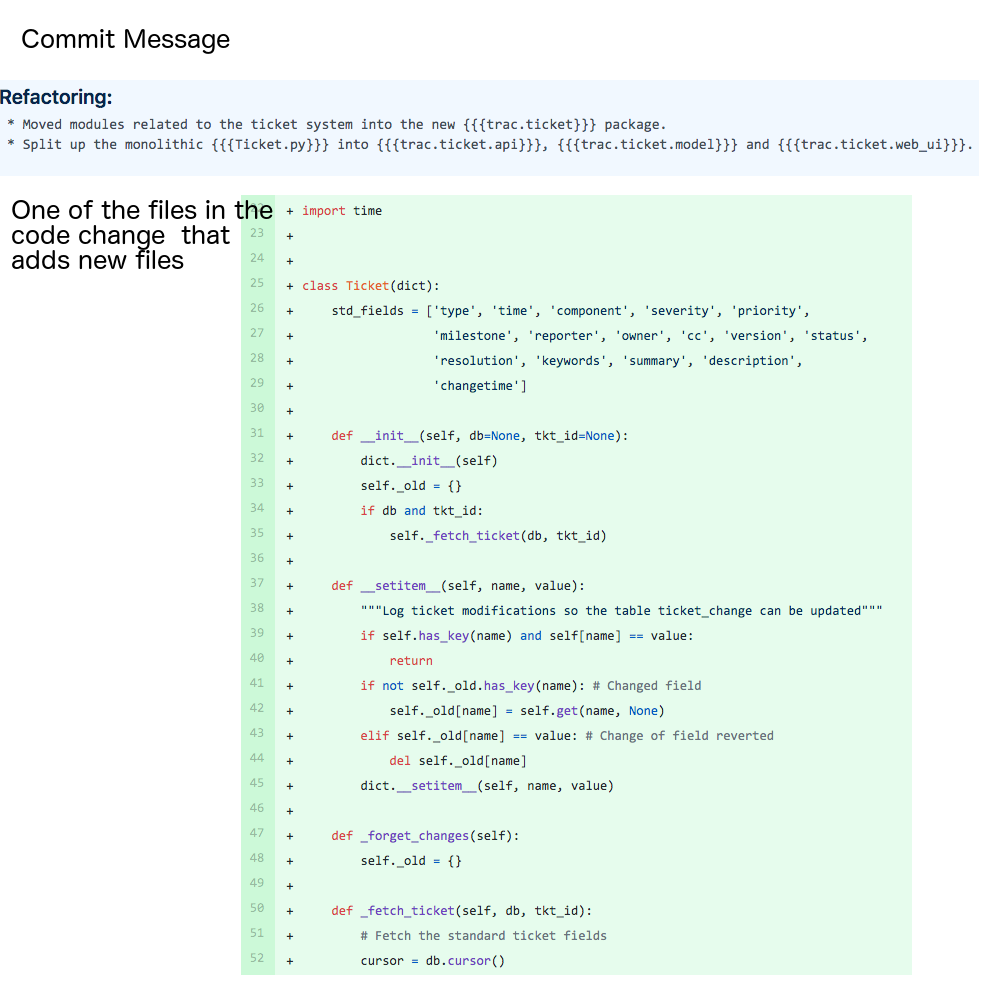}}
        \caption{\label{fig:tracrefactor} Refactor Code Change in Trac Repository.}
    \end{figure}

Refactor code changes \footnote{Commit 1fb1d3b38d32fffa50d3ff70946933f890d1d9bd} are among the defect inducing changes. This is shown in figure \ref{fig:tracrefactor}.

In such works, large code changes and new files are included. A total of 205 lines from the new file was added in the shown code change. Should this file be the defect inducing file in the code change, there is still a significant amount of inspection effort to be done by the developer.

In analyzing the code change, we compare the feature metrics to the calculated average from our previous table \ref{tab:tracstat}. Table \ref{tab:tracvs} shows the comparison.

\begin{table}[h!]
\caption{Comparison of the Code Change to Average}
\begin{center}
\begin{tabular}{c|l|l}
\textbf{Feature} & \textbf{\textit{Code Change}} & \textbf{\textit{Ave}} \\
\hline
LA & 4.67 & 6.8046 \\
LD & 4.33 & 40.6886 \\
Files Churned & 0.003 & 0.1175 \\
LT & 3 & 141.7008 \\
NS & 1 & 0.1175 \\
ND & 1 & 1.5326 \\
Entropy & 0 & 1.2246 \\
Exp & 812 & 1145.5415  \\
NUC & 1 & 2.0177 \\
\end{tabular}
\label{tab:tracvs}
\end{center}
\end{table}

This shows that the calculated feature metric for LA and LD being below the average. Making note that Kamei et al. \cite{Kamei2013} made mention that a large code changes (e.g. lines added (LA), deleted (LD), and entropy) could imply defective code change, the defect inducing code change is actually below the average for the Trac dataset. 

With the large code base of this project at 92760 LOC, locating defective code requires significant inspection effort \cite{Qiao2019}. Identifying specific lines and repairing them would need bug reports. The presence of such error means that the defect had already manifested to actual end users \cite{Qiu2020,Yan2020}.

As code changes can modify many files, we examined the number of files modified per commit and found that on average, bug inducing commits modify 7.97 files. On average, the number of different file types modified by such commits is 1.72 file types. From the analysis of the modified files, there are 39 known files that have been modified. Some files do not have file extensions and have been marked as 'no extension' instead. We have listed down the top file extensions modified on average on table \ref{tab:tracfiles}.

\begin{table}[h!]
\caption{File extensions modified in Bug inducing changes in the Trac repository}
\begin{center}
\begin{tabular}{c|c}
\textbf{File Extension} & \textbf{\textit{Average \#changed}} \\
\hline
py & 5.08 \\
html & 0.97 \\
no extension & 0.43 \\
cs & 0.39 \\
css & 0.55 \\
po & 0.19 \\
js & 0.12 \\
pot & 0.10 \\
txt & 0.09 \\
png & 0.09 \\
\end{tabular}
\label{tab:tracfiles}
\end{center}
\end{table}

\begin{figure}[h]
        \centering
        \includegraphics[width=\linewidth]
        {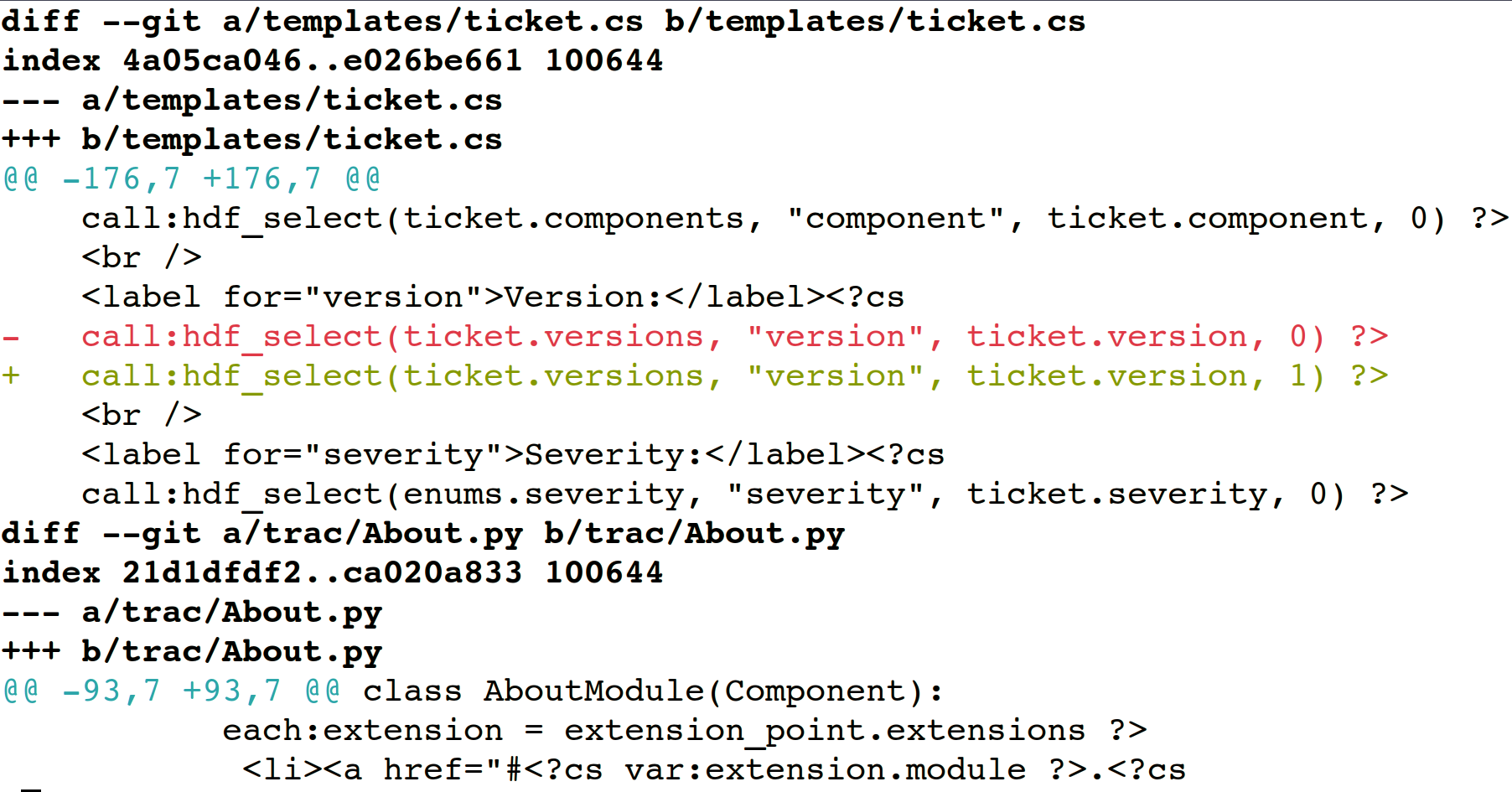}
        \caption{\label{fig:trac-diff} Sample code commit showing multiple different file types modified.}
    \end{figure}

Noting that there are potentially multiple files impacted by a specific change, identifying a commit as bug inducing does not identify \textit{which} files are defective. The developer who utilizes JIT will still need to study the commit to determine what about the code change is causing software defects. Noting that in figure \ref{fig:trac-diff} a particular code change can impact multiple file types such as .cs and .py, diagnosing the bug can be complex and time-consuming. As noted by \cite{Kamei2013}, inspecting possible defective files would take a considerable amount of time. Even if we utilize \textit{Effort-Aware} works as suggested by \cite{Kamei2013,Qiao2019}, if there is a need for the developer to study a large change, he would have to devote time and effort to understand the context of the code to determine how to fix the bug. Coarse grained approaches hinder their use from a practical standpoint \cite{Pascarella2019}.

Our results show that not only does changes in commits modify multiple files, but also changes several different file types. With our dataset including the commit ID, we are able to look into the commits marked bug inducing by the SZZ tool unlike many other available JIT datasets. One such example\footnote{Commit hash \texttt{50fc9f2a73c8e4c6bdfd92d48b80abd6e945a400}} found in our dataset is shown in figure \ref{fig:trac-diff}. The diff below illustrates two of the 46 files changed in the commit.

\begin{figure}[h]
        \centering
        \includegraphics[width=\linewidth]
        {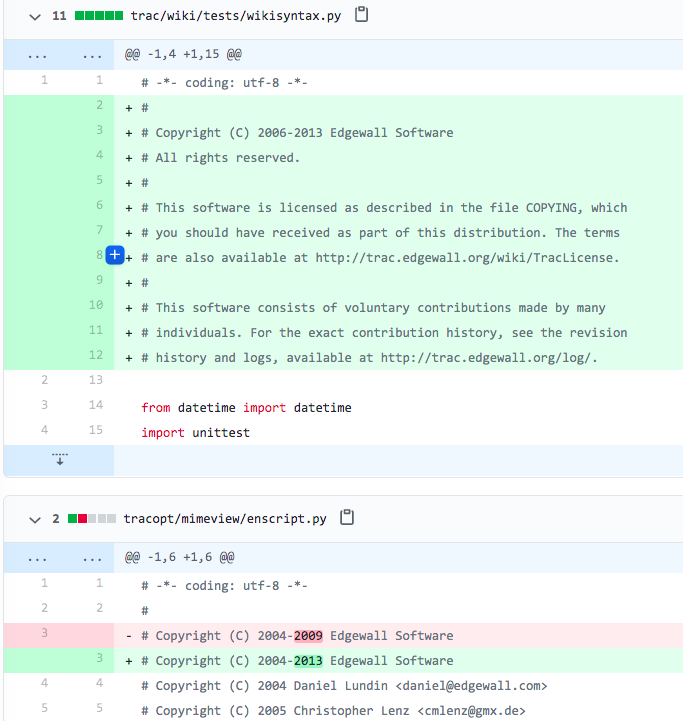}
        \caption{\label{fig:trac-copy} Sample code commit showing copyright changes.}
    \end{figure}

In analyzing other code changes, we found code changes that had large number of lines inserted and deleted. However, the code change was apparently simply adding copyright and modifying the copyright year\footnote{Commit hash \texttt{3fd4da5034e51210ec159646201f29192c2d391b}}. Figure \ref{fig:trac-copy} shows one of these code changes.

The code change itself modified 83 files and had 646 lines added and 34 line deletions. Although our experiments showed that \textit{lines of code added}, and \textit{lines of code deleted} were good features to use in predicting defective code change, further look into the metrics showed that this metric can be misleading. According to Kamei et al., large changes are more likely to be defective (e.g. lines of code added, and entropy) \cite{Kamei2013}, as the code change itself is only adding comments and updating copyright and comments, this might lead to false positive results. Also, in other works, comments have been discarded as noted in \cite{Williams2008,Hoang2019,Eghbali2020}. Comments are not typically part of the executing part of the code \cite{Eghbali2020}. The amount of inspection effort would largely go to waste for these kind of commits.

At the same time, we found large code changes that make coarse-grained JIT largely unhelpful in reducing the inspection cost. Figure shows a large code change \footnote{Commit hash \texttt{788944cebef5959a396400653801360987b27d20}} that served to significantly rewrite areas of the code.

	\begin{figure}[h]
        \centering
        \includegraphics[width=\linewidth]
        {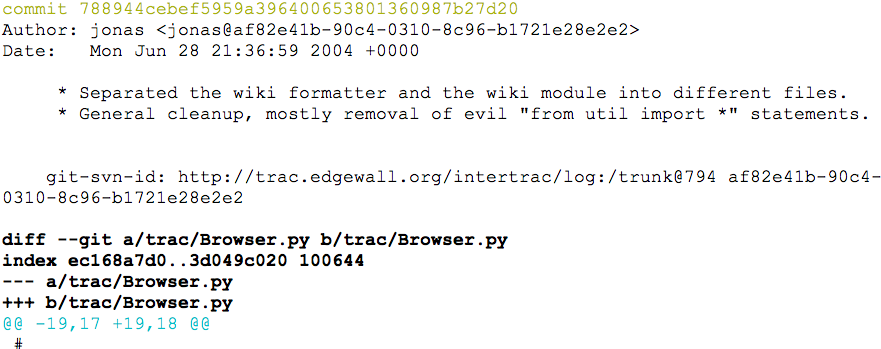}
        \caption{\label{fig:trac-buggy} Sample code commit linked with many bug fixes.}
    \end{figure}

The change itself modified 20 files with 711 lines of code added and 673 lines deleted. Interestingly to note that there has been 23 bug fixes that can be traced back to this code change \textit{over time}, several of them involved in a single file in this code change: \texttt{/trac/ticket/notification.py}. These can be called \textit{dormant bugs} where the defects are not immediately known until much later. The end result is that the defect was introduced in a much earlier version and only known to be defective in a much later version \cite{Chen2014}.

\subsection{Analyzing Qt and OpenStack}
We also looked into the two datasets provided by \cite{Hoang2019} and examine the bug inducing changes on those datasets. The summary of the datasets are shown in table \ref{tab:hoangsummary} \cite{Hoang2019}.

\begin{table}[htbp]
\caption{Summary of the datasets \cite{Hoang2019}}
\begin{center}
\begin{tabular}{c|c|c|c}
 &  &  & \textbf{\%Defect} \\
 &  &  & \textbf{induced} \\
\textbf{Project} & \textbf{Period} & \textbf{\#change} & \textbf{change} \\
\hline
Qt & 06/2011-03/2014 & 25,150 & 8\% \\
OpenStack & 11/2011-02/2014 & 12,374 & 13\% \\
\end{tabular}
\label{tab:hoangsummary}
\end{center}
\end{table}

As the list of bug inducing changes fall on different sub projects, we manually retraced the projects that were utilized for the datasets. On average, we have found that the number of files changed in the bug inducing commits in the \textit{Qt} dataset is 9.25 files and over 102 different file extensions. The highest modified file types are .cpp, .h, and .qml.

With the \textit{Openstack} dataset, on average, there are 7.19 files changed and over 33 different file extensions. The highest modified file types are .py, .json, and .xml. From these results, we can see that code changes can add or modify multiple files across file types. 

Given that on average, the number of files changed is 7.19, a developer would be expected to check the changes of these files and understand the context of such. As noted by \cite{Kamei2013}, inspection costs given a flagged defect inducing code change is important for the practical application of defect prediction.  Inspection cost and code review from coarse grained approaches also lead Pascarella et al. into identifying specific files in a code change as bug inducing \cite{Pascarella2019}.

\begin{figure}[!htb]
        \center{\includegraphics[width=\linewidth]
        {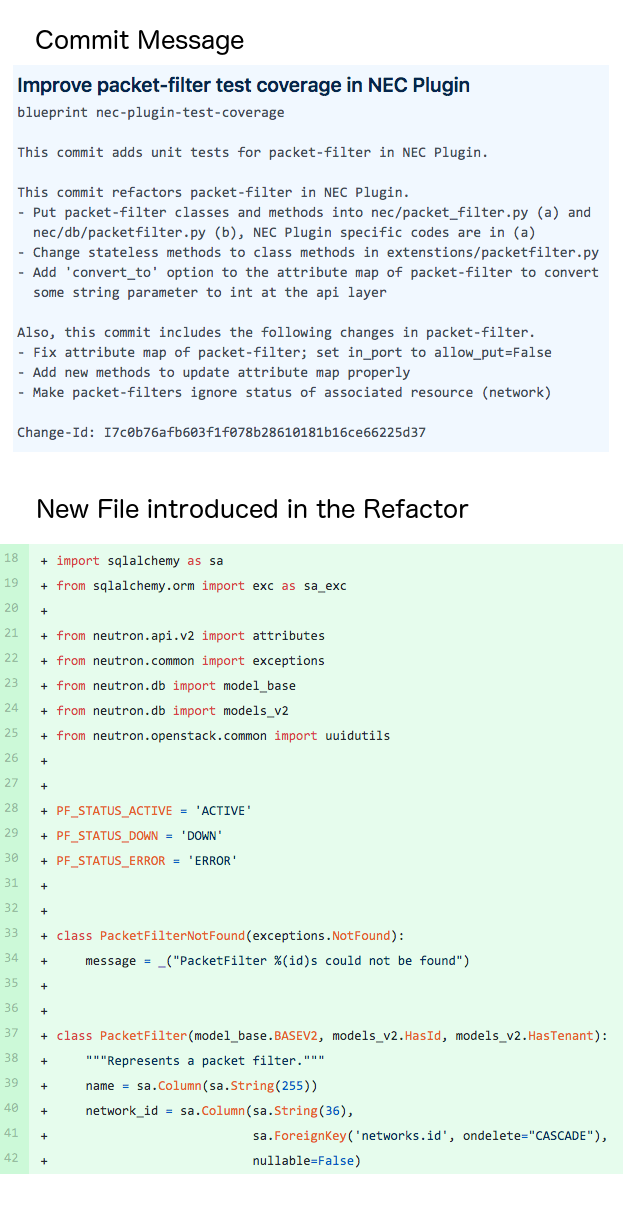}}
        \caption{\label{fig:openstackrefactor} Refactor Code Change in OpenStack.}
    \end{figure}

In analyzing the code change in the OpenStack, we also find refactor code changes that are labelled defect inducing. This is shown in figure \ref{fig:openstackrefactor}.

Similar to Trac, we find such refactoring code change to introduce new files. The number of lines added in the new file shown in figure \ref{fig:openstackrefactor} is 148.

\begin{figure}[!htb]
        \center{\includegraphics[width=\linewidth]
        {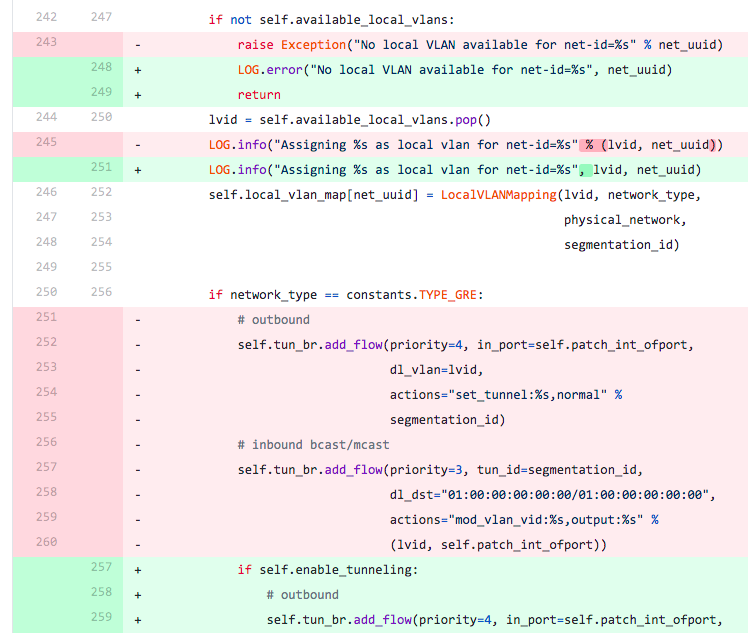}}
        \caption{\label{fig:openstacknewdiff} Modified file (ovs\_quantum\_agent.py) in a code change that adds new features added in OpenStack.}
    \end{figure}

There are defect inducing code changes]footnote{Commits such as 0061c0ce4443395a258a99c70e1c1a7d0435e84e and 00dd97c7e9a852f4abd6e8460a0d094c01d3d0ba} that add new features. For the code change\footnote{Commit 0109cc967c0a8adb9073b6e89276312cb5a95bc4}, the code change is a combination of python files and XML. Figure \ref{fig:openstacknewdiff} shows code change adding new files into the repository.

\begin{figure}[!htb]
        \center{\includegraphics[width=\linewidth]
        {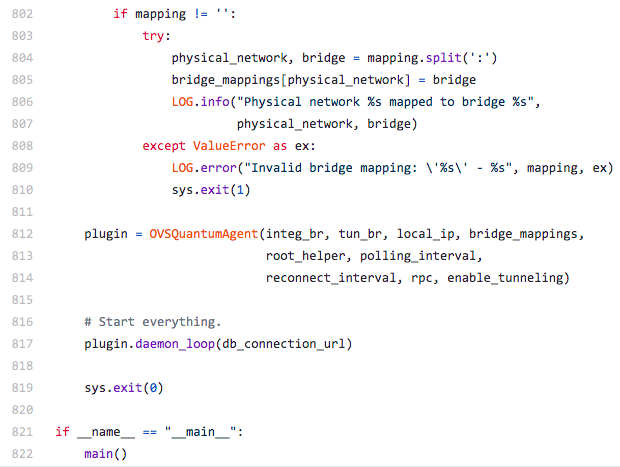}}
        \caption{\label{fig:openstackovs} End of the OpenStack Python file (ovs\_quantum\_agent.py).}
    \end{figure}

The changes are modifications of existing files, and so, there are lines removed and new lines added. This code change \footnote{Commit 00dd97c7e9a852f4abd6e8460a0d094c01d3d0ba} modifies 6 files with 163 lines added and 110 lines deleted. In marking the file within the code change as defective, the file itself is 822 lines and is shown in figure \ref{fig:openstackovs}. 

With no explainable reason for what is defective, labelling code changes and even a specific file as defect inducing is quite complex for the developer to inspect. This is consistent with the insights from \cite{Nam2018} and \cite{Dam2019}.

With the datasets provided by Hoang et al. \cite{Hoang2019} showing that bug inducing commits can span across many files and multiple file types, we believe that our findings from the trac dataset have been confirmed. Labelling a code change as defective alone may still have significant inspection effort from the developer to gauge which file or files caused the bug given a specific code change.

\section{Analysis}
Relying solely on the feature metrics alone to classify code change as defective. There have been instances when the code change is very small and yet is defect inducing. The use of semantics to capture such nuances could provide better level of detail. For example, using the code itself as input can allow the classifier to ignore code comments. This mirrors sentiments from \cite{Ghose2018}.

In addition, classifying a code change as defective may not provide enough context as to what could be defective in that commit. There is an associated inspection cost towards code changes marked as defective \cite{Kamei2013}. Although, there has been development towards \textit{Effort-Aware} researches in JIT to address this, the approach used is by sorting the changes accordingly by inspection cost \cite{Qiao2019}.

As code changes can involve many files, a code change can be \textit{partially defective} or \textit{completely defective}. Using coarse-grained JIT has inherent inaccuracies in classifying an entire code change as defective. A finer-grained approach such as files within a code change can provide better accuracy level. This is consistent with findings from \cite{Pascarella2019}.

Considering, however, the code change in figure \ref{fig:trac-buggy} that is a \textit{dormant bug} and is linked to many bug fixes, identifying the file itself as defective may not be enough. From the code changes we have shown in our analysis, it is clear that defect inducing changes are complicated. Labelling a code change as defect inducing is too simplistic a perspective. In addition, as a bug fix may end up modifying lines that were introduced by multiple code changes, the origin of a software defect could be from the combination of various code changes itself and not from a single one.

As a next step on top of automated defect prediction, being able to locate the area of the code \textit{within the code change} becomes a critical component in reducing the inspection effort of developers. Such research may need to go further than relying on the use of feature metrics with source code defects having many fine nuances that feature metrics cannot capture.

\section{Threats to Validity}
Several researches \cite{Williams2008,Davies2014,Rodriguez-Perez2018} have pointed that SZZ algorithm has limitations in determining the bug inducing changes in a repository. Our research relies on the accuracy and correctness of the existing tool provided by \cite{Borg2019}. 

Of the 2003 bug inducing changes in the \textit{Qt} dataset, we had difficulties locating 140 bug inducing hashes. We believe, however, that we have been able to demonstrate the nature of the issues with the ones we had found. We were able to trace all bug inducing hashes from the \textit{Openstack} dataset that further validates our assertions.

\section{Access to the Dataset}
The dataset is downloadable through this URL: \url{https://doi.org/10.5281/zenodo.3910002}.

\section{Conclusion}
We have presented a new dataset that can be used for JIT research that has information to extract semantic features. We also pointed out issues with current publicly available datasets. Firstly that by marking changes as defective, there is still the effort of identifying which specific file in the change as defective. Second, publicly available JIT datasets do not include the commit information that can be used to link the actual code change diffs to extract semantic features.

For future works, assembling more JIT datasets with the commit identifiers intact will allow for more diverse perspectives in progressing both JIT and semantic learning in JIT. By also looking deeper into semantic-based approaches in traditional SDP such as in \cite{Liang2019}, it is our belief that a finer-grained JIT is possible given the wealth of data that can be mined from code repositories.

\bibliographystyle{ieeetr}
\bibliography{biblio}

\end{document}